\documentclass[%
 reprint,
 amsmath,amssymb,
 aps,
]{revtex4-2}

\usepackage{graphicx}
\usepackage{dcolumn}
\usepackage{bm}

\begin{document}

\preprint{APS/123-QED}

\title{ Maxwell matter waves  }

\author{Dana Z. Anderson}

 \email{dana@coldquanta.com}
\affiliation{ColdQuanta Inc., 3030 Sterling Circle, Boulder, Colorado 80301 USA and Department of Physics and JILA, University of Colorado, Boulder, Colorado, 80309-0440, USA}

\date{\today}

\preprint{APS/123-QED}

\begin{abstract}
Maxwell matter waves emerge from a perspective, complementary to de Broglie's, that matter is fundamentally a wave phenomenon whose particle aspects are revealed by quantum mechanics.  Their quantum mechanical description is derived through the introduction of a matter vector potential, having frequency $\omega_0$, to Schrodinger's equation for a massive particle.  Maxwell matter waves are then seen to be coherent excitations of a single-mode of the matter-wave field. In the classical regime, their mechanics is captured by a matter analog of Maxwell's equations for the electromagnetic field. As such, Maxwell matter waves enable a spectrum of systems that have useful optical analogs, such as resonant matter-wave interferometric sensors and matter-wave parametric oscillators. These waves are associated with a wavelength that is tied to the drive frequency $\omega_0$ rather than to the massive particle's energy, as is ordinarily the case with de Broglie matter waves. As a result, simple interferometric measurements lead to different outcomes for the two types of waves. While their apparent departure from de Broglie character is surprising, Maxwell matter waves are wholly consistent with quantum mechanics.  
 \end{abstract}
 
 \date{\today}
\maketitle

\section{Introduction}

The term ``matter wave'' is generally associated with the wave properties exhibited by massive particles having momentum $p$, which are characterized by the de Broglie  wavelength relationship $\lambda = h/p$, where $h$ is Planck's constant \cite{zubairy.2020}.  Whereas a wave is a spatially extended entity, the classical view of a particle ascribes to it a well-defined position.  Quantum mechanics dictates that the localization of a particle to with a span of $\delta x$ requires a spread of particle momenta $\delta p$ consistent with Heisenberg's uncertainty principle, $\delta x \delta p \geq \hbar/2$, where $\hbar$ is the reduced Planck's constant.  A particle said to be propagating would furthermore have momentum characterized by $\left|\delta p/p \right| \ll 1$.   

Here we will use the term ``de Broglie matter wave,'' or simply ``de Broglie wave,'' to refer to the wave characteristics associated with a flux of identical massive particles having fixed energy $E_{\rm{v}}$ and a given spectral width. A de Broglie wave can  transmit power, do useful work, and be utilized in sensing applications, for example.  In particular, however, the spectral components of a de Broglie matter wave have no specific relationship to each other: each particle is independent of the other particles comprising the entire flux.  Because the individual particles are associated with a spectral width, we understand that an interaction such as a collision will generally result in a distribution of energy outcomes. 

A view of matter that is different from the customary de Broglie wave becomes apparent as one seeks to identify states in which interactions lead to a definite energy change.  In particular, given a free particle having vacuum energy eigenvalue $E_{\rm{v}}$ (and therefore it is not a localized particle), due to an interaction it subsequently bares energy $E_{\rm{v}}'=\hbar(\omega_{\rm{v}}+\omega_0)$ that is also an energy eigenvalue. This leads  to the notion of a single-mode matter-wave field identified by its frequency $\omega_0$.  In the presence of this field a particle can absorb an additional $\hbar \omega_0$ amount of energy, or return to its original vacuum state by releasing that amount of energy back into the field.  We will show that in a classical limit this field is associated with a pair of propagating oscillatory fields:
\begin{eqnarray}\label{Eq:OscillatoryFields}
     & \label{Eq:MatterEField} \textbf{F}\left(x,t\right)=\textbf{F}_0 \sin\left(k x - \omega_0  t\right), \\
     &\label{Eq:MatterBField} \textbf{ G}\left(x,t\right)=\textbf{G}_0 \sin\left(k x - \omega_0  t\right),
\end{eqnarray}
which are analogous to the electric and magnetic fields $\textbf E $ and $\textbf B$ associated with waves, say, produced by an electronic microwave oscillator, though they do not propagate at the speed of light. As unintuitive as electric and magnetic matter analogs might be, they are inescapable consequences of the single-mode formalism.  In any case, the analogy between electromagnetic and matter waves proves to be generally fruitful. We thus introduce the term ``Maxwell matter wave,'' or sometimes simply ``Maxwell wave,'' to refer to classical states of the matter-wave field.  

We originally introduced Maxwell matter waves in \cite{Anderson.2021}, there referred to as ``classically coherent matter waves,'' as the emission generated by an atomtronic transistor oscillator \cite{Caliga.2012nlo, Caliga.2016i7o, Dinardo.2018}. For the sake of this work we do not explain how a matter-wave oscillator functions; we simply assume that such an instrument exists.  Instead we wish to establish that Maxwell matter waves are consistent with quantum mechanics and discuss some of their elementary aspects.  

The oscillator involves atomic ensembles having a temperature $T \lesssim \hbar \omega_0 /k_B$, where $k_B$ is Boltzmann's constant, so that the energy associated with the emitted particles exceeds their thermal energy. Such low temperatures, on the order of $50$nK in the experimental work of \cite{Dinardo.2018},  is also a regime in which the de Broglie aspects of matter will be manifest.  One might say that de Broglie waves are the essence of the wave-particle duality of quantum mechanics in which one thinks classically in terms of particles that prove to exhibit wave-like behavior under the laws of quantum mechanics. Yet the Maxwell matter waves expressed by the fields Eqs. (\ref{Eq:MatterEField}) and (\ref{Eq:MatterBField}) are clearly classical and clearly more wave-like than they are particle-like.  Maxwell waves provide a complementary view of matter within the wave-particle duality: classical waves that prove to exhibit particle-like behavior under the laws of quantum mechanics.  We will reflect below on the two views of matter in the context of Bose-Einstein condensation. 

Louis de Broglie tendered his hypothesis regarding the wave properties of matter in 1924 — nearly 100 years ago — and it was validated by the experiments of Thomson \cite{THOMSON.1927} and of Davisson and Germer \cite{davisson.1927b} in 1927. These early diffraction experiments and many validations since have placed de Broglie's momentum-wavelength relationship on firm ground. We shall establish here that the massive particles associated with Maxwell matter waves do not exhibit the expected de Broglie relationship.  The conclusion might seem to contradict de Broglie's hypothesis, if not quantum mechanics more fundamentally, but it does not. It is merely surprising.  We unravel the apparent inconsistency in the Remarks section. 

While the distinction between de Broglie and Maxwell matter waves is meaningful, the practical utility of Maxwell matter waves provides our primary motivation here.  In that context our purpose is to lay some ground work by cultivating the analogy between Maxwell matter waves and electromagnetic waves.  By doing so one makes available to the matter-wave domain the tremendously extensive and powerful set of tools that have been developed for the electromagnetics domain.  We shall apply such tools to an example inertial sensing application. Drawing out the analogy also enables a deeper understanding of the nature of Maxwell matter waves and their distinction from de Broglie waves.  

Our approach starts with a classical description of matter in terms of particles, transitions to a quantum description which interprets particles as waves, then comes full circle to a classical description of matter in terms of coherent state excitations of a matter-wave field.

\section{Classical Particle Description}
The classical-to-quantum transition begins with a classical derivation of the Hamiltonian, so we will start by considering a particle having mass $m$ propagating in the $x$-direction.  Charged masses repel or attract according to whether they have same or opposite charge signs.  Neutral identical particles also typically repel or attract; in the ultracold domain they interact as though they are spherical.  Let us leverage the analogy by subjecting our particle to a vector potential; the conjecture that it leads to a valid single-mode picture is to be validated later by the outcome.  We choose a vector potential that points as well as travels in the $x$-direction:
\begin{equation}\label{Eq:VectorPotential}
    \textbf{A} = \hat{x}A_0\cos(k x - \omega_0 t).
\end{equation}
The Lagrangian $\mathcal L$ for a particle is given by the difference of its kinetic and potential energies, $T$ and $V$:
\begin{equation}
    \mathcal L = T - V,
\end{equation}
in which the kinetic energy is:
\begin{equation}
    T= \frac{1}{2}m \dot{x}^2, 
\end{equation}
and the potential energy:
\begin{equation}
    V = m\left(\frac{\omega_0}{k}-\dot x \right) A_0\cos(k x-\omega_0 t).
\end{equation}
The Lagrangian is determined from the energies: 
\begin{equation}
    \mathcal L 
     =\frac{1}{2}m \dot{x}^2 + m(\dot x-\frac{\omega_0}{k})  {A}_0 \cos(k x - \omega_0 t).
\end{equation}
The canonical momentum $p$ is determined from the Lagrangian:
\begin{equation}
    p=\frac{\partial \mathcal L}{\partial \dot{x}}=m \dot{x} + mA_0 \cos(k x-\omega_0 t).
\end{equation}
The Hamiltonian is then:
\begin{equation}
      H=p\dot{x}-\mathcal L = \frac{1}{2}m \dot{x}^2 + \frac{m\omega_0}{k} A_0\cos(k x-\omega_0 t),
\end{equation}
or:
\begin{equation}
\begin{split}
    H &= \frac{1}{2m} \left|p-m A_0\cos(k x-\omega_0 t)\right|^2   \\
    &+ \frac{m\omega_0}{k} A_0\cos(k x-\omega_0 t).
\end{split}
\end{equation}
Now Hamilton's equations of motion for the position and momentum give,
\begin{equation}
    \dot{x}=\frac{\partial H}{\partial p}=\frac{p}{m}-A_0 \cos (k x-\omega_0 t),
\end{equation}
\begin{equation}
\begin{split}
    \dot{p}=-\frac{\partial H}{\partial x}&=(m\omega_0-p k) A_0 \sin (k x-\omega_0 t)\\
    &+2k mA_0^2\cos(k x-\omega_0 t)\sin(k x-\omega_0 t).
\end{split}
\end{equation}
Note that in the limit of small vector potential, $A_0 \ll p/m$, the canonical momentum is constant provided:
\begin{equation}\label{Eq:GroupVelocity}
    v_{\rm{v}}\equiv \frac{\omega_0}{k}=\frac{p}{m}.
\end{equation}
We comment that the atomtronic oscillator inherently fulfills this condition by the nature of its dynamics. Therefore the system's canonical momentum is nearly time independent, $p=p_{\rm{v}}$, and the system Hamiltonian is simply that of a free particle and of the field separately:
\begin{equation}\label{Eq:Hamiltonian}
    H=\frac{p_{\rm{v}}^2}{2m}+\frac{1}{2}mA_0^2\cos^2(k x-\omega_0 t).
\end{equation}
The particle's kinetic momentum $P$, however, oscillates in time and space around its mean value:
\begin{equation}
    P\equiv m \dot{x} = p_{\rm{v}}-mA_0\cos(k x-\omega_0 t).
\end{equation}

\section{Quantum Mechanical Description}
With the Hamiltonian in hand, the transition from the classical to the quantum regime is relatively straightforward, replacing Hamilton's equations for position and momentum with Schrodinger's for the wave function $\psi$, $\hat{\mathcal H}\psi = E \psi$.  The Hamiltonian, Eq. (\ref{Eq:Hamiltonian}), looks to have the form of a particle subject to a moving optical lattice, the solution to which can be expressed in terms of Bloch functions.  The difference lies in the dispersion relation from Eq. (\ref{Eq:GroupVelocity}), $k=\omega_0/v_{\rm{v}}$. That is, the ``lattice'' is tied to the particle, rather than to an independent frame of reference.  

In absence of the vector potential, the time-independent Schrodinger equation admits the fundamental de Broglie matter wave: the single-mode plane wave solutions for a wave function —solutions that conceptually occupy all of space.  
\begin{equation}\label{Eq:SingleMode}
    \psi = \frac{1}{\sqrt{N}}e^{i(k_{\rm{v}} x - \omega_{\rm{v}}t)},
\end{equation}
for which, as above, $E_{\rm{v}}=p_{\rm{v}}^2/2m=\hbar \omega_{\rm{v}}$, $k_{\rm{v}} = p_{\rm{v}}/\hbar$ and $N$ is an appropriate normalization factor.  

The vector potential field must also be quantized, which can be done using the same method as for electromagnetic field \cite{Scully.1997}. In so doing we introduce of a set of creation and annihilation operators $\{\hat{a}^{\dagger} (\omega_0), \hat{a}(\omega_0) \}$.  These operators act on a single mode of the matter-wave field, raising or lowering the energy by $E_{\rm{m}}=\hbar \omega_0$. Notably, they do not create and destroy massive particles. The packets of energy, dubbed ``matterons'' in Ref. \cite{Anderson.2021}, are the matter-wave analogs of the photon.  The vector potential field, Eq.  (\ref{Eq:VectorPotential}), is a coherent state, that is, an eigenstate of the annihilation operator $\hat{a}$.  The state is conventionally written as $|\alpha\rangle$:
\begin{equation}
    \hat{a}|\alpha\rangle = \alpha |\alpha\rangle .
\end{equation}
As such, the vector field does not have a well-defined number of matterons or energy,
\begin{equation}\label{Eq:MeanEnergy}
    \langle E_{\alpha} \rangle = \left(|\alpha|^2+ \frac{1}{2} \right) \hbar \omega_0;
\end{equation}
however, the excited matterwave is associated with exactly one matteron, so that each particle itself has a well-defined vacuum energy, $E_{\rm{v}}'=\hbar (\omega_{\rm{v}}+\omega_0)$.  Such definite energy was our objective.  Notice that the energy associated with the particle's canonical momentum $p_{\rm{v}}$ is the same as that of the particle in absence of the field, while the additional $\hbar \omega_0$ can be viewed as internal energy associated with the dipole motion of the particle, which itself is associated with momentum having magnitude $\hbar |k|$.  

As our focus is on their classical properties, here we defer important aspects concerning the zero-point energy $\hbar \omega_0/2$ and spontaneous phenomena associated with the matter-wave field to a later section. For related reasons, we also sidestep questions about multi-matteron excitation associated with a single massive particle.  

The coherent state character of the vector potential is embossed on the state of the matterwave field; it seems thus fitting to refer the result as a ``coherent excitation of the matter-wave field''. This brings us full circle and enables us to treat an ensemble of particles classically as we do in the following sections, but as a wave rather than as a localized propagating entity.  It is interesting to note, in fact, that a localized, particle-like, entity can be constructed either as a superposition of de Broglie waves associated with a spectrum of energies $\hbar\delta \omega_{\rm{v}}$, or as a superposition of Maxwell matter waves having a spectrum of excitation energies $\hbar \delta \omega_0$. In the latter case, however, energy uncertainly is reflected through the uncertainty in the number of particles.

\section{Matter wave Maxwell analogs}
The atomtronic oscillator emits a constant flux of particles we identify as a Maxwell matter wave. We derived the mechanics with the introduction of a vector potential, the matter analog of the vector potential utilized in electromagnetics.  Evidently, though perhaps less intuitively, we can equally well describe the behavior in terms of matter analogs to electric and magnetic fields.  While we have borrowed the electromagnetic nomenclature with the symbol ``$\textbf {A}$'' for the vector potential, we use different symbols for the electric and magnetic field analogs to underscore the fact that matterwaves do not involve electric charges and currents:
\begin{equation}\label{Eq:Efield}
    \textbf{F} =- \frac{\partial \textbf{A}}{\partial t},
\end{equation}
\begin{equation}\label{Eq:Bfield}
    \textbf{G}=\nabla \times \textbf{A} ,
\end{equation}
from which we substantiate our assertion of Eq's (\ref{Eq:MatterEField}) and (\ref{Eq:MatterBField}). These fields propagate at the group velocity of the particle.  Indeed, it seems that the fields are intimately tied to the particles, and thus we can describe the physics either in terms of the fields or in terms of the particles.  The latter context is very much like the fields associated with the carriers in an electrical circuit; namely, instead of the fields we can equivalently work with the matter analogs of electric potential and electric current.  We will thus refer to the analogs as ``potential'' and ``current:''
\begin{eqnarray}\label{Eq:OscillatoryWaves}
     & \label{Eq:PotentialWavess} {\mathcal V}\left(x,t\right)={\mathcal V}_0 \cos\left(k x - \omega_0  t\right), \\
     &\label{Eq:CurrentWave} \mathcal I\left(x,t\right)={\mathcal I}_0 \cos\left(k x - \omega_0  t\right).
\end{eqnarray}

Generally the fields are suited to a matter-wave optics context while the $\mathcal{V}$-$\mathcal{I}$ treatment is particularly appropriate in an atomtronic circuit context. In experiments, particles are typically manipulated and confined in wave-guide structures.  In this regime the physics is indeed much like that of microwave circuitry.  Writing equations in terms of potentials and currents allows one to use the familiar mathematical tools for analyzing matter-wave circuits such as transmission line theory \cite{Gonzalez.1997}.  More generally, Kirchhoff's laws and the concepts of resistance and reactance, are all at play in atomtronic circuit analysis. The SI units associated with matter-wave quantities can simply be found by substituting ``kilogram'', or ``kg'', for every occurrence of ``Coulomb'', or ``C''. Thus electric potential as J/C becomes matter potential J/kg and electrical capacitance as C$^2$/J becomes matter capacitance kg$^2$/J. 

The potential and current amplitudes are related through a characteristic impedance ${Z}$\cite{Gonzalez.1997}: 
\begin{equation}
	{\mathcal V}_0 =   Z{\mathcal I}_0, 
\end{equation}
where
\begin{equation}
	Z=n^2 \frac{ \hbar} {m^2}\equiv n^2 Z_0 .
\end{equation}
Here $n$ plays the role of a refractive index:
\begin{equation}
	n \equiv \sqrt{\frac{ \omega_{\rm{0}} }{\omega_{\rm{v}}}},
\end{equation}
where recall that $E_{\rm{v}}\equiv \hbar \omega_{\rm{v}}$ is the vacuum energy of individual particles associated with the wave. While it is uncommon in optics to have an refractive index less than unity, here it is easily conceivable to have $1/n\gg1$. 

As we here look to highlight the classical nature of Maxwell matter waves, we take the natural impedance $Z_0$ to be the fundamental entity describing wave properties rather than Planck's constant and substitute $\hbar=m^2 Z_0$. As evident from Eq.  (\ref{Eq:GroupVelocity}) the waves propagate at the group velocity of a deBroglie wave:
\begin{equation}\label{wavevelocity}
	v_{a}=\frac{1}{n}\sqrt {2 m \omega_0 Z_0 } \equiv \frac{v_0}{n} ,
\end{equation}
and are thus associated with the wave number:
\begin{equation}\label{Eq:wavenumber}
	k=n\sqrt{\frac{ \omega_0  }{2 m Z_0}}\equiv n k_0	.
\end{equation}

Though superfluous for our present purposes, we can complete the matter-wave field analogy with electromagnetic fields by defining matter-wave permeability and permittivity:
\begin{equation}\label{Eq:permeability}
\begin{split}
     & \mu_0 \rightarrow \upsilon_0 \equiv \sqrt{\frac{ Z_0}{2 m \omega_0}}, \\
     & \mu \rightarrow \upsilon = n^3\upsilon_0 .
\end{split}
\end{equation}
\begin{equation}\label{Eq:CurrentField} 
\begin{split}
     &\epsilon_0 \rightarrow \xi_0 \equiv \frac{1}{Z_0\sqrt{2 m \omega_0 Z_0}}, \\
     &\epsilon \rightarrow \xi =  \xi_0/n.
 \end{split}
\end{equation}
Table \ref{Table:1} provides a list of analogs between matter-wave quantities and electromagnetic quantities.

The matter-wave fields satisfy:
\begin{equation}\label{Eq:MWMaxwell}
     \nabla \times \textbf{F} = -\frac{\partial \textbf{G}}{\partial t},  
\end{equation}
\begin{equation}
    \nabla \times \textbf{G} = \upsilon \xi \frac{\partial \textbf{F}}{\partial t} ,
\end{equation}
from which it follows they also satisfy a familiar wave equation:
\begin{equation}
    \frac{\partial^2 \textbf{F}}{\partial t^2}=\frac{1}{\upsilon \xi}\frac{\partial^2 \textbf{F}}{\partial x^2},
\end{equation}
and similarly for $\textbf{G}$ (as well as $\mathcal V$ and $\mathcal I$). We recognize that the wave's phase velocity $1/\sqrt{\upsilon \xi}$ is the particle's group velocity.  Note, too, that the Maxwell wave phase and group velocities are equal.  

\begin{table}[ht]
    \centering
    \setlength{\tabcolsep}{9pt}
    \begin{tabular}{ l c c l }
         EM Element & EM & MW &  SI Units\\ 
        \hline
        Vector potential & $\textbf {A}$ & $ \textbf {A}$ & $\rm{m} \cdot \rm{s}^{-1}$\\
        Electric field & $\textbf {E}$ & $\textbf {F}$ & $\rm{m} \cdot \rm{s}^{-2}$ \\ 
        Magnetic field & $\textbf {B}$ & $\textbf {G}$ & $\rm{s}^{-1}$ \\
        Electric potential & $\mathcal V$ & $\mathcal V$ & $\rm{m}^2 \cdot \rm{s}^{-2}$ \\
        Current & $\mathcal I$ & $\mathcal I$ & $\rm{kg} \cdot \rm{s}^{-1}$ \\
        Impedance & Z & Z & $\rm{m}^2 \cdot \rm{kg}^{-1} \cdot \rm{s}^{-1}$ \\
        Capacitance & C & C & $\rm{kg}\cdot \rm{s}^2 \cdot \rm{m}^{-2} $ \\
        Inductance & L & L & $\rm{m}^2 \cdot \rm{kg}^{-1} $ \\
        Permittivity & $\epsilon$ & $\xi$ & $\rm{kg} \cdot \rm{s}^2 \cdot \rm{m}^{-3}  $ \\
        Permeability & $\mu$ & $\upsilon$ & $\rm{m} \cdot \rm{kg}^{-1}  $ \\
        \hline 
    \end{tabular}
    \caption{Matter-wave (MW) analogs to electromagnetic (EM) quantities and corresponding SI units}
    \label{Table:1}
\end{table}

It is useful to appreciate that the current wave, Eq.  (\ref{Eq:CurrentWave}) does not imply a time-varying particle flux, any more, say, than the electromagnetic field of a laser beam implies a time-varying photon flux.  The wave amplitudes are determined from Eq.  (\ref{Eq:MeanEnergy}), or equivalently through the (experimentally measurable) particle flux $I_{\rm{v}}$ given in units of particles/s:
\begin{equation}\label{Eq:particleflux}
	{\mathcal I}_0 =\frac{m}{n}\sqrt{2\omega_0 I_{\rm{v}}}.
\end{equation}

\begin{figure}[t]
\includegraphics[width=\columnwidth]{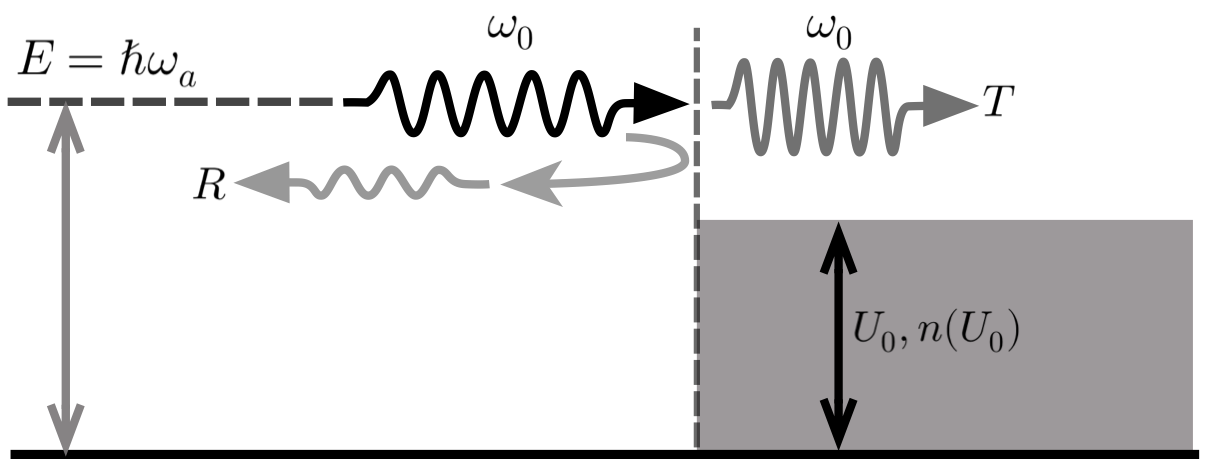}
\caption{\label{fig:Step} The reflection and transmission coefficients $R$ ad $T$ of a Maxwell matter wave from a step change in refractive index yields the same result as that of a de Broglie wave from a step potential.  The oscillation frequency $\omega_0$ as well as the particle energy $E=\hbar \omega_{\rm{v}}$ remain constant while the wavelength decreases in the higher index region.    }
\end{figure}

\section{Properties}
Let us contrast the nature of the Maxwell wave with the wave associated with a particle having energy $E_{\rm{v}}$.  The latter has wavenumber:
\begin{equation}\label{Eq:deBroglieWavenumber}
    k_{\rm{v}}=\sqrt{\frac{2m\omega_{\rm{v}}}{\hbar}}=\sqrt{\frac{2\omega_{\rm{v}}}{m Z_0}}=\frac{2}{n} k_0.
\end{equation}
The distinction is clear, since the de Broglie wavenumber increases with increasing particle energy, yet decreases for the Maxwell matter wave case, Eqn. (\ref{Eq:wavenumber}).  While the conclusion might seem to contradict de Broglie's hypothesis that a particle's wavelength is inversely proportional to its momentum, we consider the elementary quantum mechanics problem of reflection from a step potential of height $U$, as illustrated in FIG. \ref{fig:Step}. The refractive index for the Maxwell waves may be generalized to:
\begin{equation}\label{Eq:GeneralizedIndex}
	n(U) \equiv \sqrt{\frac{ \omega_{\rm{0}} }{\omega_{\rm{v}}}}\left(1-\frac{U}{m^2\omega_{\rm{v}}Z_0}\right)^{-1/2},
\end{equation}
whereas that for the de Broglie waves is:
\begin{equation}
    \tilde{n}(U)=\frac{1}{n(U)}.
\end{equation}
Amplitude reflection and transmission of a wave from a potential step is given by a well-known pair of formulas \cite{Hecht.2017}; for the Maxwell matter wave:
\begin{equation}
\begin{split}
     & r=\frac{n_1-n_2}{n_1+n_2}, \\
     & t=\frac{2n_1}{n_1+n_2},
 \end{split}
\end{equation}
for which we have taken $n_1=n(0)$ and $n_2=n(U_0)$.  Using the same formulas for the de Broglie wave we find:
\begin{equation}
\begin{split}
     & \tilde{r}=-r ,\\
     & \tilde{t}=\frac{n_2}{n_1} t,
 \end{split}
\end{equation}
The transmitted and reflected particle flux in the two cases are identical:
\begin{equation}
\begin{split}
     & R\equiv |r|^2=\tilde{R} ,\\
     & T\equiv |t|^2=\tilde{T}.
 \end{split}
\end{equation}
We have indicated that the Maxwell matter fields are associated with a constant flux $I_{\rm{v}}$ of mono-energetic particles, thus the Maxwell and de Broglie treatments of the interface are consistent.  The conclusions are very different, however, when propagation involves multiple paths for which the coherence properties of the wave come into play.

\begin{figure}[t]
\includegraphics[width=\columnwidth]{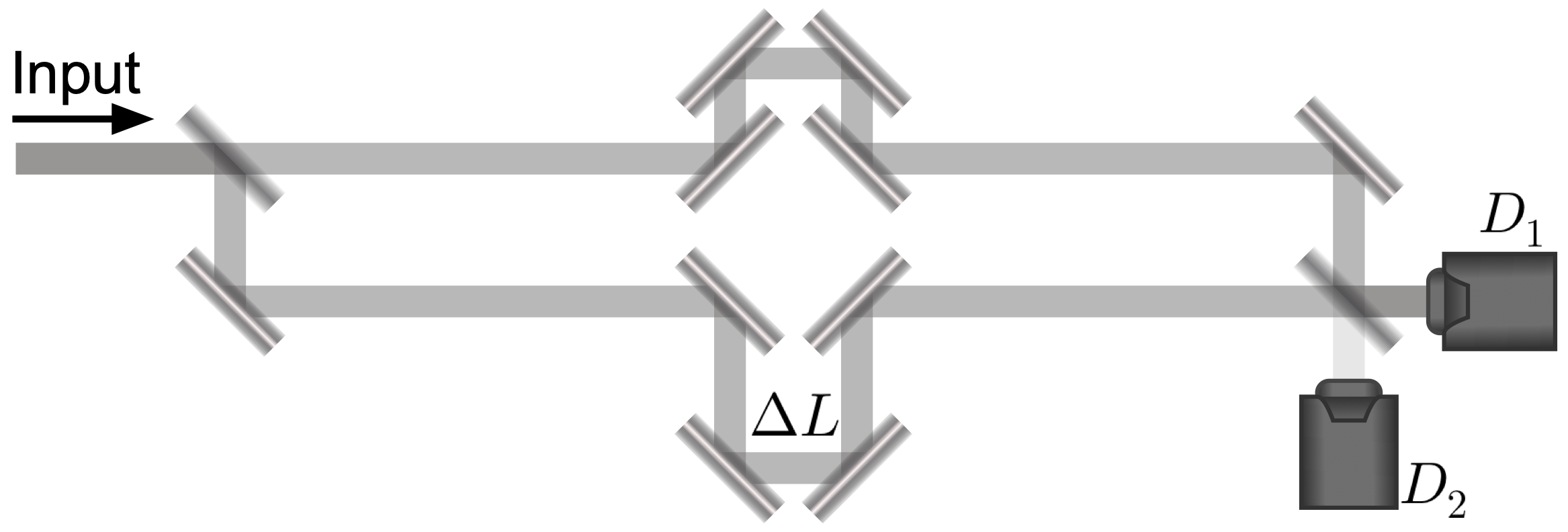}
\caption{\label{fig:MachZehnder} Interference of matter waves in a Mach-Zehnder interferometer will reveal the distinction between de Broglie and Maxwell matter waves.  The ``fringe'' spacing, i.e. the change required in the path length difference $\Delta L$ to cause the detected particle flux to change from a maximum to zero in the two cases will differ by a factor of $n^2/2$ }.
\end{figure}

Consider the Mach-Zehnder interferometer of FIG. \ref{fig:MachZehnder}.  For atomic particles it is conceptually straightforward to implement mirrors and beamsplitters utilizing A.C. Stark shifts induced with narrowly-focused knife-shaped laser beams tuned to the blue side of an atomic resonance. One arm of our interferometer incorporates an adjustable path length and there is a particle detector at the output of each port.  The properties of the interferometer are well known \cite{Hecht.2017}:  For identical arm lengths one port will be ``dark'' meaning no particles are detected while the other will be ``bright'' meaning the entire input atom flux will be detected.  The question is: by how much does the path length difference need to be changed to reverse the dark and bright ports?  Taking the case that $\omega_{\rm{v}} \gg \omega_0$ the de Broglie result would indicate a much smaller path length difference than the Maxwell wave result, by a factor of $n^2/2$.  The two calculations yield different predictions for the particle count as a function of path length difference, so Maxwell and de Broglie waves are definitively distinct.

\section{Resonant Matter-Wave Interferometry}
\begin{figure} [t] 
\includegraphics[width=\columnwidth]{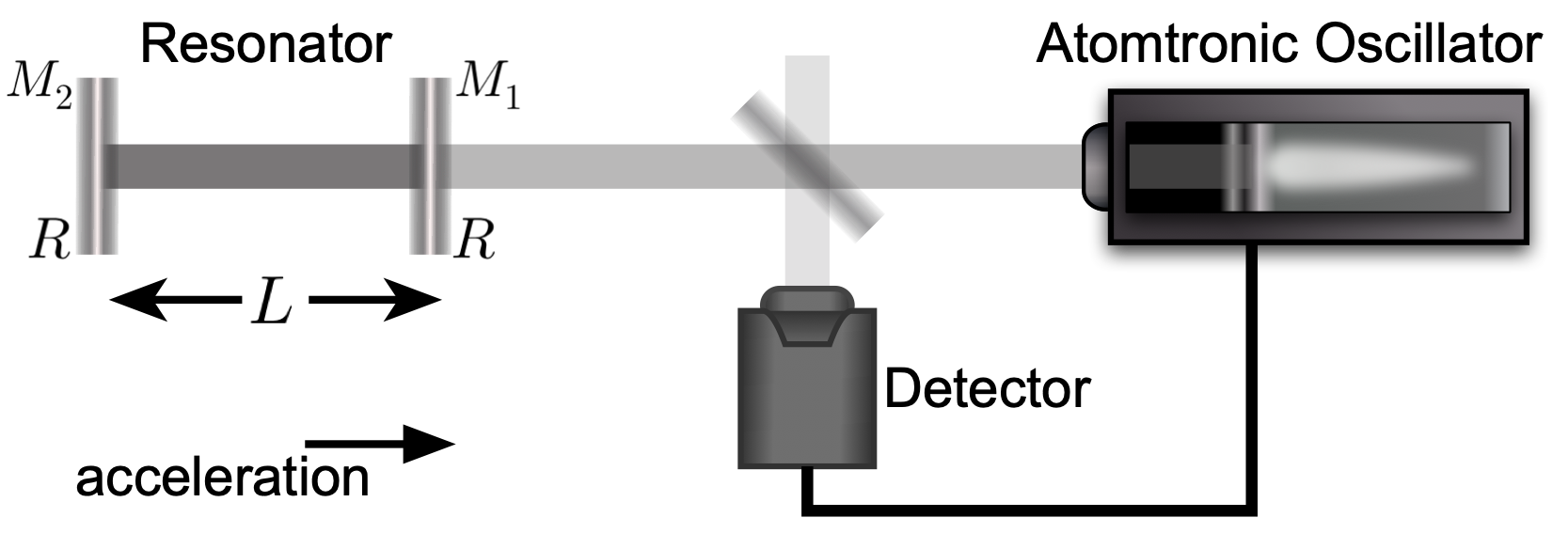}
\caption{\label{fig:FabryPerot} Notional scheme of a resonant matterwave interferometer for sensing acceleration is comprised of a pair of matterwave ``mirrors''.  The drive frequency of an atomtronic oscillator is servo-controlled to maintain resonance using a matter-wave version of Pound-Drever-Hall feedback arrangement \cite{Black.2001}.  The presence of an acceleration induces a resonant frequency shift.    }
\end{figure}

The temporal coherence of the Maxwell matter waves accommodates the matter-wave analog of the optical resonator, and in particular enables resonant matter-wave interferometry.  FIG. \ref{fig:FabryPerot} illustrates a notional system in which a matter wave is injected into a resonator formed with a pair of matter-wave reflectors having reflectively $R$. To simplify the discussion we suppose that the flux is sufficiently low that particle interactions can be ignored.  Feedback to the matter-wave generator adjusts the oscillator frequency to maintain its frequency on resonance \cite{Black.2001}.  The treatment of the matter-wave resonator follows identically that for the optical case: the mirrors are separated by a distance $L$ chosen to be an integer number $N$ of half-wavelengths of the Maxwell wave, $k L=n k_0L = \pi N$, to produce a resonating matter-wave interferometer. Converting the resonant wavelength to a frequency, 
\begin{equation}
    \omega_{\rm{N}}=N\frac{\pi v_{\rm{v}}}{L}.
\end{equation}
A measure of the quality of a resonator is given by the finesse \cite{Hecht.2017}:
\begin{equation}
    F=\pi \frac{\sqrt{R}}{1-R},
\end{equation}
from which one can determine a resonator linewidth:
\begin{equation}
    \delta \omega = \frac{\pi v_{\rm{v}}}{ L F}.
\end{equation}

Consider now the the presence of an inertial force, let us say causing an acceleration $a$ along the resonator axis as shown in FIG. \ref{fig:FabryPerot}: in the frame of the resonator the particle energy varies during propagation in the resonator. The acceleration can be treated as a spatially-dependent refractive index that induces an apparent change in the path length:
\begin{equation}
   n(x)=n\left( 1 + \frac{ax}{m\omega_{\rm{v}}Z_0}\right)^{-1/2}.
\end{equation}
Mimicking its optical analog we refer to the optical path length:
\begin{equation}
    L_{\rm{eff}}=n \int_{0}^{L} \left( 1 + \frac{ax}{m\omega_{\rm{v}}Z_0}\right)^{-1/2} dx .
\end{equation}
Taking the acceleration to be small:
\begin{equation}
    L_{\rm{eff}} \simeq n L\left(1-\frac{L}{4m\omega_{\rm{v}}Z_0}a\right).
\end{equation}
The acceleration $a$ can be determined by locking the oscillator frequency to on particular resonant mode, $\omega_0 =\omega_{\rm{N}}$ and  measuring its frequency shift $\Delta \omega_{\rm{N}}\equiv \omega_{\rm{N}}^\prime - \omega_{\rm{N}}$,
\begin{equation}
    \Delta \omega_{\rm{N}} = \kappa a,
\end{equation}
for which the scale factor
\begin{equation}
    \kappa=\frac{\pi N }{2 v_{\rm{v}}} .
\end{equation}
Associated with the finite linewidth of the resonator will be a characteristic acceleration resolution of the interferometer:
\begin{equation}
    a_{\rm{res}} = \frac{2\pi }{ F  }\frac{v_{\rm{v}}^3}{\omega_0 L^2} .
\end{equation}
The key result here is that the acceleration signal corresponds to a frequency shift of the resonator, and the acceleration resolution is determined by the resonator finesse, which can be large. Setting, for example, $v_{\rm{v}} = 0.01$m/s, $L=0.01$m, $\omega_0 = 2 \pi \times 1$kHz, and $F=100$, yields $a_{\rm{res}}=10^{-7}$m/s${}^2$. Given the small size, this compares favorably with cold atom interferometers when provided with even a modest atom flux, say, of $10^3$ particles/second assuming particle shot-noise limited performance \cite{Berman.1997}.

The matter-wave resonator, like its optical counterpart illuminated with appropriate laser light, relies on the temporal coherence of the Maxwell matter wave.  Resonance can viewed as a compounding of constructive interference, versus the lack of it, between the wave that reflects off of the input mirror and that which has been transmitted from the built up, stored matter wave, bouncing between the two mirrors. Resonant systems find many uses in the optical and more generally electromagnetic domain.  We expect the same can be true for resonant matter-wave systems.

\section{The Atom Analog of the Laser}
Bose-Einstein condensation takes place when a sufficiently high density of particles is brought to temperatures near absolute zero \cite{Pethick.2002}.   The Bose-Einstein condensate (BEC) itself is an ensemble of particles in thermal equilibrium. Quantum mechanically the BEC is a number state, meaning a state having a fixed number particles.  At zero temperature the particles comprising a BEC are described by a common localized wave function having a well-defined energy.  While the coherence of a BEC was debated at the time \cite{Castin.1997}, Ketterle and collaborators' unequivocal demonstration of matter-wave interference laid to rest any doubt about the quantum coherence ascribable to a condensate \cite{Andrews.1997}. In principle one knows everything about the position and momentum of the particles, and also the ensemble of particles as a whole, that the laws of quantum mechanics allow. In this respect the BEC is the epitome of the deBroglie perspective in which matter is comprised of particles whose wave-like character is imposed by quantum mechanics. 

From the time of its first demonstration in 1995 \cite{Anderson.1995bec,Davis.1995} the BEC was hailed as the atom analog of the laser \cite{Kleppner1997}. A laser is a nonthermal equilibrium system that belongs to the general class of driven oscillators, conceptually a close cousin to electronic transistor oscillators: both produce classically coherent electromagnetic fields whose mechanics is governed by Maxwell's equations. Indeed the particle aspects of light articulated by Planck and by Einstein marked the birth of quantum theory.  Whereas a BEC is comprised of a localized ensemble of particles having a definite number, a coherent state is nonlocal, does not have a well-defined number of particles or energy, nor a particle wave function from which one can determine the probability of finding a particle at a given position at a given time \cite{Scully.1997}.  The atomtronic transistor circuit that produces Maxwell matter waves is also a driven oscillator and its emission is also a coherent state, but of matter rather than light, with its particle-like aspects attributed to matterons as the analog to photons.  So while there are aspects of a BEC that resemble the laser, we now understand that observations of de Broglie matter-wave interference of BECs is distinct from the laser-like interference of Maxwell matter waves.

\section{Interactions}
That one can leverage the formalism afforded by Maxwell equations is a compelling aspect of a matter-as-wave classical perspective.  One still faces the duality of quantum mechanics, however, and in the context of matter interactions the particle view of matter appears to provide a more natural picture. Yet the matter-as-wave picture was motivated in the first place by the search for a representation in which particles could interact and yet remain in energy eigenstates; the search led us to the notion of a matteron as the analog of the photon. In contrast to photons, matterons can evidently interact. 

Consider first the interaction between counter-propagating but otherwise identical Maxwell matter waves.  The resonator discussed in the previous section serves as a illustrative model in which to consider such a case.   Elastic events for which the energy carried by the waves remains unchanged can be treated in a mean-field context.  The energy density associated with the particle flux $I_{\rm{v}}$ of the forward-propagating wave is:
\begin{equation}
    u = \hbar k  \frac{I_{\rm{v}}}{\mathcal{A}} ,
\end{equation}
where $\mathcal{A}$ is an effective cross-sectional area of the matter wave.  For the simple case in which the particle interaction is characterized by a scattering length $a_{\rm{s}}$, particles associated with the backward-traveling wave experience a mean-field interaction energy:
\begin{equation}
	H_{\rm{int}} =  8\pi\frac{a_{\rm{s}} u}{k_0^2} .
\end{equation}
The impact of the interaction can be cast in terms of an refractive index shift $\delta n$, following from  Eq.  (\ref{Eq:GeneralizedIndex}):
\begin{equation}
    \delta n \simeq  4 \pi n^4\frac{a_{\rm{s}}}{k_0}\frac{I_{\rm{v}}}{A}.
\end{equation}
The significance for resonant interferometry is that the index shift causes a corresponding shift of the resonant frequency.  

In addition to elastic interactions, one can also anticipate spontaneous parametric interactions arising from matteron exchange between the counter-propagating beams. The simplest spontaneous interaction generates two new waves, one having an increased energy by $\hbar \omega_0$ and the other decreased by the same amount, and correspondingly a momentum increase of $\hbar k$ for the one wave and decrease by the same amount for the other.  The question of momentum conservation can be addressed through the refractive indexes $n_+, n_-$ corresponding to the waves having acquired or lost a matteron, respectively:
\begin{equation}\label{Eq:Interaction}
    n_{\pm} =  \frac{n}{\sqrt{1 \pm n^2}} \simeq n \left( 1\mp  \frac{1}{2} n^2 \right),
\end{equation}
where we have taken the case $\omega_{\rm {v}} \gg \omega_0$. The change in the canonical momentum can be found from Eq. (\ref{Eq:deBroglieWavenumber}):
\begin{equation}
    \Delta p_{\rm{v}} = 2 \left(\frac{1}{n_+}-\frac{1}{n_-}\right) \hbar k_0 \simeq 2 \hbar  k.
\end{equation}
Thus we see that the change in canonical momentum compensates for the change in momentum due to the matteron exchange. 

The example of spontaneous interactions can highlight several important aspects of the Maxwell matter wave formalism.  It is clear, for example, that matteron number is not conserved:  In the above example one matteron was removed from each initial wave to create two new waves. Yet each of the initial waves is an eigenstate of an annihilation operator, i.e. a coherent state, and so the mean number of matterons does not change. Moreover, the process does not involve the creation or annihilation of mass - the state of matter is merely altered.  

Finally and importantly, the presence of parametric interactions suggests the possibility of parametric amplifiers and oscillators, which are systems that are familiar particularly within  optics \cite{Saleh.1991}.  Indeed, this brings us full circle in providing some insight toward the construction of a matter-wave oscillator, the output of which is a Maxwell matter wave.

\section{Remarks}
We have developed a framework in which matter is viewed classically in terms of waves rather than particles. Within this framework, matter can be treated using analogs to Maxwell's equations and surrounding formalism. The presence of matter analogs of electric and magnetic fields is non-intuitive: Maxwell unified the two seemly different phenomena of electricity and magnetism to be recognized as one phenomenon, that of electromagnetism.  Maxwell's electromagnetic waves have an existence of their own, yet viewed in the context of matter, electromagnetism is simply a description of the interaction among charged massive particles.  Reflecting on our earlier comments, neutral atoms such as $^{87}$Rb repel each other, as do same-sign charges.  The repulsive forces can be cast in terms of chemical potential energy, which becomes large as the particle density becomes high, as in a BEC \cite{Pethick.2002}. The atomtronic oscillator \cite{Dinardo.2018,Anderson.2021} incorporates a harmonic atomic potential having frequency $\omega_0$ and develops a time- and space- varying potential in a self-organized fashion (as one might also expect from a parametric oscillator): this is the physical origin of the forces that underlie the introduction of the vector potential, Eq. (\ref{Eq:VectorPotential}). The analogy between electromagnetic and matter interactions is a faithful one in many respects: given a matter vector potential, the electric field and magnetic field analogs mathematically follow. 

The alternative to de Broglie's view of matter sought to identify states in which particle interactions preserved their energy-eigenstate character. This led to the notion of  single-modes of the matter-wave field. The previous section illustrated that coherent state excitations of these modes lead to a picture in which the particle interactions do preserve their eigenstate character, even though the waves themselves are not energy eigenstates.     The aspect that is fundamentally new to the theoretical foundation is the presence of a matter-wave field, associated with which are matterons as massless particles having energy $E_{\rm{m}}=\hbar \omega_0$ and momentum $p_{\rm{m}}= \hbar k$.  Implicit also in the formalism is the presence of matter awaiting excitation, so to speak, since the matterons do not create or annihilate mass.  

The distinction between Maxwell and de Broglie matter waves associated with the same particle energy is highlighted by the substantially different outcomes of interferometric measurements.  This difference can be easily understood starting from the classical mechanics treatment: the particle center of mass is oscillating along the direction of propagation at frequency $\omega_0$.  Surely the center of mass must arrive from each of the two arms at the same time at the output port to produce constructive interference.  The quantum mechanical particle has no center of mass \textit{per se}, but its momentum nevertheless oscillates with the same period. Constructive interference can only take place when the interferometer arm length difference corresponds to a propagation time difference equal to an integral number of those periods.   

Maxwell matter waves have significant practical implications.  Among them is the possibility of using the waves in conjunction with resonant structures, as already described.  More subtle is the role of interference in simple matter-wave analogs of optical elements, such as a diffraction grating. Such subtleties and other aspects of Maxwell matter waves can be addressed with the well-developed framework of electromagnetics, appropriately translated to the matter-wave domain. 

Matteron interaction lends both a complexity and a richness to the wave picture of matter, especially since the strength of interactions can be controlled.  In the case of atoms one can utilize electromagnetic fields to tune interactions, utilizing for example Feshbach resonances \cite{Chin.2010} and Rydberg interactions \cite{Shaffer.2018} to suppress or enhance the interaction strength. Matteron interactions highlight the particle character of Maxwell matter waves and potentially makes uniquely quantum-mechanical aspects such as squeezing and entanglement substantially accessible and controllable.

\begin{acknowledgments}
The author is grateful to D. Gu{\'e}ry-Odelin, S. Du, and M. Saffman for valuable feedback and discussion.
\end{acknowledgments}

\bibliography{MMW}

\end{document}